\documentstyle[12pt]{article}
\newcommand{\gs}{$g_{6}\;$}
\newcommand{\gsb}{$\overline {g_{6}}\;$}

\begin{document}
\title{Lattice on Lattice in Two Dimensions}
\author{Ronald Dickman and Eugene M. Chudnovsky\\
        Department of Physics and Astronomy\\
        Herbert H. Lehman College, CUNY\\
        Bedford Park Boulevard West\\
        Bronx New York 10468-1589\\}
\maketitle
\begin{abstract}
We study a harmonic triangular lattice,
which relaxes in the presence of a
weak, short-wavelength
periodic potential.  Monte Carlo simulations reveal
that the elastic lattice has only short-ranged positional
correlations, despite the absence of defects
in either lattice.
Long-range orientational order, however, persists in the
presence of the background.  Our results provide an
alternative explanation for the hexatic glass
phase observed in high-temperature superconductors,
magnetic bubble lattices, and charge density waves in
semiconductors.
\end{abstract}
\vspace{2em}


\newpage
Vortex lattices in
superconductors \cite{murray,blatt}, magnetic bubble lattices
in ferromagnetic films \cite{sesh}, charge
density waves \cite{dai} and Wigner
crystals in semiconductors \cite{andrei}, are examples of
elastic lattices which exist in a
periodic potential
having a much smaller period, and
a different
symmetry. In these systems, it is reasonable to treat the atomic
lattice as static, since its elastic moduli
are much
larger than those of the large-scale lattice.
It is widely believed that when the
interaction between the two lattices is weak,
crystalline order in
the elastic lattice may be destroyed as a result of
defects in either of the two lattices.
The main purpose of this letter is to show that, rather
surprisingly, {\em no disorder} in the atomic lattice is required
to destroy translational long-range order in a
defect-free elastic
lattice.
We perform Monte Carlo simulations of a
two dimensional harmonic
``lattice on lattice" system, in
which the background lattice is
static and defect-free, in order to learn the
effect such a background has on translational
and orientational order.  We find that after
the elastic lattice is cooled slowly
to temperature zero, translational
correlations are short-ranged, but that
long-range orientational order persists.  It follows
that results of recent experiments \cite{murray,sesh,dai}
on elastic lattices subject to a background can be
understood {\em without appealing to defects in these
quenched disorder in the underlying atomic lattice.}

A fundamental
observation of Halperin and Nelson \cite{halnel}
is that in disordered two dimensional lattices,
orientational order is more robust than
translational order. Consequently, there is a
{\em hexatic} phase with short range
translational order, but extended correlations in
the orientation of local crystallographic
axes.
Two kinds of hexatic phase have been studied to
date. The first
appears in the theory of two-dimensional melting.
As is known \cite{ll1,mermin}, in two
dimensions, true long-range translational order exists only at zero
temperature. At low temperatures translational
correlations exhibit power-law decay, while
orientational order is long-range.
Above the Kosterlitz-Thouless temperature \cite{kt},
dislocation pairs dissociate
into free dislocations \cite{kt,berez}, and the solid
melts into a hexatic phase \cite{halnel} in
which translational correlations decay exponentially with distance, while
orientational correlations decay as a power law. Above the Halperin-Nelson
temperature, dislocations (which are equivalent to pairs of disclinations)
dissociate into free disclinations, which transforms the hexatic phase into a
true liquid, with both translational and orientational
correlations decaying exponentially.

A different kind of hexatic phase appears in the
presence of quenched disorder at zero temperature,
as in random pinning of such structures as vortex lattices, magnetic bubble
lattices, and charge density waves, by defects in the underlying atomic
lattice.
Larkin \cite{larkin,blatt} demonstrated that
quenched disorder, no matter how weak, destroys
long-range translational order. The Imry-Ma
theorem \cite{imryma} generalizes this observation to assert
that any long-range order described
by a continuous-symmetry order parameter is destroyed by arbitrarily weak
static random forces, below five dimensions. It was noticed later, however,
that orientational order is more stable with  respect to quenched
randomness than is translational order \cite{emc1}.
At low temperatures it decays as
a power law, leading to a solid phase called a hexatic glass \cite{emc2}.
The existence of this phase has been firmly established in vortex
lattices of high-temperature superconductors
\cite{murray}, in magnetic bubble lattices \cite{sesh},
and in charge density lattices in
semiconductors \cite{dai}.  There is also reason
to believe that at low temperatures, rare gas monolayers
on solid substrates \cite{nagler,greiser},
and two-dimensional Wigner crystals in semiconducting
heterojunctions \cite{andrei}, also exist in the
hexatic glass phase.

There is a vast literature on computer simulations of two-dimensional
lattices in the presence of thermal disorder
\cite{strand}. Simulations of the effect
of quenched randomness have been less systematic and are limited to magnetic
systems \cite{serota,dieny,rdemc}. In this letter we
report a numerical study of translational
and orientational correlations in a triangular lattice of $10^{4}$
sites, subject to a weak periodic potential representing a
quadratic lattice of much higher spatial frequency.

We consider a two dimensional triangular
net of ``particles" coupled by a
harmonic, nearest-neighbour interaction. Let ${\bf a}_{1}, ... , {\bf a}_{6}$
be the basis vectors for the lattice, i.e., ${\bf a}_{1}=(a,\,0),\;
{\bf a}_{2}=(a/2,\,{\sqrt{3}}a/2)$, etc. The position of each particle in the
ideal lattice is denoted by ${\bf r}_{i}$, with
$i=1,2,...,N$; ($N$ is the total number of sites). The
position of the i-th particle in the deformed lattice is
${\bf {\Upsilon}}_{i}={\bf {\Upsilon}}({\bf r}_{i})$,
and the corresponding displacement is
$ {\bf u}_{i}={\bf u}({\bf r}_{i})=
{\bf {\Upsilon}}({\bf r}_{i})-{\bf r}_{i} $.
The potential energy of the lattice is purely harmonic:
\begin{equation}
U_{latt}=\frac{1}{2}\sum_{i=1}^{N}\sum_{j=1}^{6}\{{\bf a}_{j}{\cdot}
[{\bf u}({\bf r}_{i}+{\bf a}_{j})-{\bf u}({\bf r}_{i})]\}^{2}\;\;\;.
\label{harm}
\end{equation}
(All quantities are dimensionless in our
formulation; $a=1$ in
simulations.)
The particles
are subject to a periodic
potential having the symmetry of a square lattice,
\begin{equation}
V(x,y)=S\,cos(kx)\,cos(ky)\;\;\;,
\label{bknd}
\end{equation}
where $2{\pi}/k\,<<\,a$. The interaction energy then becomes
\begin{equation}
U_{int}=S\sum_{i=1}^{N}cos\{k[x_{i}+u_{x}({\bf r}_{i})]\}\,
cos\{k[y_{i}+u_{y}({\bf r}_{i})]\}\;\;\;.
\label{uint}
\end{equation}

The translational correlation function is \cite{halnel}
\begin{equation}
g_{\bf G}({\bf R})=<e^{\{i{\bf G}{\cdot}[{\bf u}({\bf r}_{i}+{\bf R})-
{\bf u}({\bf r}_{i})]\}}>=<cos{\{{\bf G}{\cdot}[{\bf u}({\bf r}_{i}+{\bf R})-
{\bf u}({\bf r}_{i})]\}}>\;\;\;,
\label{trcorr}
\end{equation}
where ${\bf G}$ is a reciprocal lattice vector and the average is taken over
all sites. The angle brackets denote
an equilibrium average, in the case of finite-temperature
systems.  For the lattice on lattice problem at
zero temperature, they denote an average over initial
configurations (at a certain temperature) as well as an
average over noise during the ensuing cooling process.
(Thus we are not particularly interested in the ground
state of the system, but rather in typical configurations
into which the lattice becomes frozen.)
The orientational correlation function is defined
as \cite{halnel}
\begin{equation}
g_{6}(r)=<e^{i6[{\theta}({\bf r}_{i})-{\theta}({\bf r}_{j})]}>=
<cos{\{6[{\theta}({\bf r}_{i})-{\theta}({\bf r}_{j})]\}} >\;\;\;,
\label{orcorr}
\end{equation}
where $r=|r_{i}-r_{j}|$ and ${\theta}_{i}$ is the angle between
${\bf {\Upsilon}}({\bf r}_{i}+{\bf a}_{1})-{\bf {\Upsilon}}({\bf r}_{i})$
and the x-axis.

We briefly review some exact results
of elastic theory for the continuum
limit of our discrete model,
Eq. (\ref{harm}), in the absence of any background. In the continuum limit,
the triangular lattice is described by
an isotropic elastic tensor involving two constants \cite{llel},
\begin{equation}
U_{latt}=\frac{1}{2}\int\,d^{2}r\,(2{\mu}u_{{\alpha}{\beta}}u_{\alpha \beta} +
{\lambda}u_{{\gamma}{\gamma}}^{2})\;\;\;,
\label{ucont}
\end{equation}
where
\begin{equation}
u_{{\alpha}{\beta}}=\frac{1}{2}\left[\frac{{\partial}u_{\alpha}}
{{\partial}r_{\beta}}+\frac{{\partial}u_{\beta}}{{\partial}r_{\alpha}}
\right]\;\;\;
\label{strain}
\end{equation}
is the strain tensor, ${\alpha},{\beta}=\{x,y\}$.
(In Eq (\ref{ucont}) repeated indices are summed over.)
Comparison with the long-wavelength expansion of Eq (\ref{harm}) reveals that
$ \mu = \lambda =\frac{\sqrt{3}}{2}a^{2} $.
Analysis of the continuum version of our model along the lines of
Ref.\cite{halnel} yields for the translational correlation function at the
first Bragg point, $G=4{\pi}/a{\sqrt{3}}$,
\begin{equation}
g_{\bf G}^{T}=1/R^{\eta}\;\;\;,
\label{power}
\end{equation}
with
\begin{equation}
{\eta}=\frac{32{\pi}}{9{\sqrt{3}}}T\;\;\;.
\label{eta}
\end{equation}
A similar analysis of the orientational correlation function yields
\cite{halnel}
\begin{equation}
g_{6}=exp\left(-\frac{9{\Lambda}^{2}}{8{\pi}{\mu}}T\right)\;\;\;,
\end{equation}
where ${\Lambda}$ is the ultraviolet cutoff. For ${\Lambda}{\sim}2{\pi}/a$
this gives
\begin{equation}
ln(g_{6})\,{\sim}\,-3{\sqrt{3}}{\pi}T\;\;\;.
\label{lng6}
\end{equation}
These results will be used to test
our numerical method.

Our main goal is to compute translational and orientational
correlation functions in the presence
of the underlying incommensurate, rapidly oscillating, periodic potential,
Eq(\ref{bknd}).
This potential represents a highly non-linear contribution to the energy,
and is not readily amenable to
analytical treatment. One might suppose that it mimics
some of the effects of a {\em random}
background, in that the potentials affecting
sites separated by more than a few lattice spacings are essentially
uncorrelated. It may be useful, therefore, to compare our results with the
predictions of a model with
quenched randomness, {\em i.e.}, an
interaction of the form
\begin{equation}
U_{int}=\sum_{i}{\bf f}({\bf r}_{i}){\cdot}{\bf u}({\bf r}_{i})\;\;\;,
\label{force}
\end{equation}
where f$_{\alpha}({\bf r}_{i})$ is a Gaussian random force
with standard deviation $\sigma$. Continuum
elastic theory \cite{emc2} predicts that translational correlations are
approximately Gaussian in form,
$ g_{\bf G}^{f}({\bf R}) \approx exp[- (R/\xi)^2]$,
with
\begin{equation}
\xi \approx \sqrt{27/14 \pi} \sigma ^{-1}
\label{xirf}
\end{equation}
at the first Bragg point.
Orientational correlations are predicted
to decay as a power law,
$g_{6}^{f}(R)=1/R^{{\eta}_{6}}$,
with
${\eta}_{6}=\frac{3}{2{\pi}} \sigma^{2}$.
It is of course not clear that
static disorder can be used
to model the effects of
the incomensurate periodic background.  In any case
one might suppose that a random {\em potential} would
be more pertinent than the simple (but solvable)
random force model.  Although there is no reason to
expect a Gaussian form for the translational
correlation function in the presence
of a random potential, the scaling of the correlation
length, $\xi \propto 1/\sigma$ can be derived on the
basis of a more general Imry-Ma type argument.

In our simulations we considered lattices
with both periodic and open
boundaries.
The former systems were rhomboid in form, with $M \leq 100$ lattice sites
to a side; the latter were hexagonal, with $M \leq 60$.
In the open-boundary systems we compute correlation functions
over the interior only, i.e., for sites at least $M/2$
distant from the edge.
We computed the translational correlation function,
Eq. (\ref{trcorr}), for two reciprocal lattice vectors,
both of magnitude $4 \pi/\sqrt {3}$,
making angles of $30^o $
and $90^o$, respectively, with {\bf R}, which lies along a principal
lattice axis.  In what follows we report
on $g_{\bf G}$ only for the former orientation;
the results
for the other orientation are very similar.

The model is simulated
{\it via} the usual Metropolis procedure
for a system in equilibrium at temperature T.
At each step, a particle is selected at random and subjected to a
trial displacement, in a random direction, and with a random magnitude,
uniform on [0,D].  (We take $D = max [2T, 0.05]$.)
The move is accepted if the total change in energy $\Delta E \leq 0$.
If  $\Delta E $ is positive the new position is accepted only with
probability $e^{-\Delta E/T}$.  In studies of lattices subject to a
background potential, the simulations begin at a temperature $T \simeq S$, the
strength of the background.  Then the temperature is gradually
reduced, typically
over a series of ten steps, each consisting of about $10^3$
lattice updates, before the system
is permitted to relax at temperature zero for several thousand lattice updates.
(A ``lattice update" represents
one trial displacement, on average, per particle.)
For background potential strength $S = 0.2$, for example,
we used the temperature sequence: $T = $ 0.2, 0.13, 0.09,
0.06, 0.04, 0.25, 0.17, 0.12, 0.008, 0.005, each for
1400 lattice updates, followed by 3500 lattice updates at
$T = 0$.
The number of lattice updates per temperature step was
sufficient that by the end of that step, the energy had
reached an essentially constant value.   We have also
performed some preliminary studies using a continuously
decreasing temperature.
Correlation functions are computed in the final state.

As a check on our simulation method, we first
consider the
well-understood case of a harmonic lattice at finite temperature,
in the absence of any background potential.
Periodic systems with $M = 100$ ($10^{4}$ sites),
and open systems with $M= 60$,
($N= 10621 $ sites), were studied in five independent
realizations at  temperatures
ranging from 0.02 to 0.2.
The results were independent of the boundary
conditions.
The mean elastic energy $\langle U_{latt} \rangle = NT$
as required by equipartition.
Fig. 1 shows correlation functions for
temperatures 0.05 and 0.2.; the initial
decay in the translational correlation function
is well-described by a power law, $g_{{\bf G}} \approx
A r^{-\eta}$, as predicted by
elastic theory, Eq(\ref{power}).  Power law decay of
the positional correlations is observed at all of the
temperatures studied, with
$\eta /T = 6.6 \pm 0.1 $, while theory gives 6.45 for
this ratio.
The orientational correlation $g_6$, also shown in Fig. 1,
attains an essentially constant value \gsb after a few lattice spacings.
\gsb is an exponentially decreasing function of temperature,
as predicted by elastic theory,
with $- \ln \overline{g_6}/T = 17.3$.
Eq(\ref{lng6}) yields 16.3 for this ratio, but this
prediction depends on the value
one takes for the cutoff $\Lambda$,
which is not precisely known.
Thus the agreement between
theory and simulation for the thermal case is quite satisfactory.

We studied systems subject to the potential $V(x,y)$,
with the coupling
$S$ ranging from 0.02 to 0.5,
averaging in each case over 10 independent realizations.
A snapshot of a typical zero-temperature configuration
(Fig. 2),   shows that while local positions are clearly
disordered, the crystallographic directions are well-
defined and globally correlated.
Examples of correlation functions obtained for systems
with open boundaries
are shown in Fig. 3.
The correlations decay fastest at small r, which is incompatible
with the Gaussian form predicted for the random force
model.  In fact, the positional correlations are
reasonably well-described by a simple exponential.
The associated correlation length does not appear to follow
any simple scaling with $S$, but if we identify $S$
with $\sigma$ in the random force model,
then Eq (\ref{xirf}) yields the correct order of magnitude:
$ \xi_{\rm rf} = 39.2$, $\xi_{\rm sim} = 24.7$, for $S = 0.02$;
$ \xi_{\rm rf} = 7.8$, $\xi_{\rm sim} = 10.7$, for $S = 0.1$.
As in the thermal case, \gs attains a steady value after a few
lattice spacings.  While the random force model
predicts power-law decay of orientational correlations,
we note that, the decay is
very slow, {\em i.e.}, $\eta_6 \approx 3 S^2 /2\pi
\approx 0.1$ for $S = 0.5$, the largest value considered.
Comparison of the correlation lengths and of \gsb for the two wave numbers
studied ($k=39.984$, and $k=20.0$), reveals no significant dependence upon k.
We note that for the larger wavenumber,
the potential energy per particle is $-S$, even for the smallest
$S$-values studied, while for $k=20$
saturation does not occur until $S \geq 0.1$.  This indicates that for
the larger wavenumber there is a potential minimum sufficiently near a
particle's equilibrium position that
all particles can reach such a
minimum.

We also studied harmonic lattices with {\it periodic}
boundaries, subject to a
background potential, as above, in systems of length $M=100$,
with $k= 39.984$.
Positional correlations (Fig. 4) now decay
more slowly than an exponential, but faster than a power law; for $r$ not too
large they are well-represented by a ``stretched exponential," form,
$g(r) = exp[-ar^{\beta}]$.  The decay rate $a$ is proportional to
the potential strength S, while the exponent $\beta$ {\em decreases }
with $S$ (for $S=0.05$, $\beta = 0.71$; for
$S=0.2$, $\beta = 0.55$).  Repeating the $S=0.1$ study on
a larger system ($M=160$) yielded essentially identical
results.
As in the obc simulations, orientational correlations
are essentially independent
of $r$ beyond a few lattice spacings; \gsb is in fact independent of
the boundary conditions.  We have also found stretched
exponential decay of translational correlations in systems
with {\em open} boundaries, when the temperature is decreased
continuously.  The correlation length is comparable to that
obtained for step-wise cooling, and long-range orientational correlations
persist.  (A systematic investigation of cooling rate
effects will be published
elsewhere \cite{unpub}.)  We stress that while the precise form of the
translational correlation function depends on the cooling schedule and
boundary conditions,  the appearance of hexatic order
is independent of such details.

We propose the following scenario for the loss of
translational order.
As it is cooled in the periodic background potential,
certain regions of the elastic lattice become
frozen.  Since the background is incommensurate
with the elastic lattice, local order is reduced
in the frozen regions.  Moreover, these regions
are well-separated, so that particle positions
are not correlated between them.  This does not
preclude the presence of long-range orientational order.
Thus we find that disorder typical of a hexatic glass may
emerge in an elastic lattice,
cooled in the presence of an
underlying atomic lattice, even in the absence of
quenched disorder.  This provides
an alternative explanation
for recent experiments on such
elastic lattices, and suggests
that the hexatic phase is the
generic low temperature state of such systems.

We are grateful to the Department of Energy
for support under Grant No. DE-FG02-93ER45487.
Simulations employed the facilities of the University Computing
Center of the City University of New York.

\pagebreak

\noindent{\bf Figure Captions}

\vspace{1em}
\noindent Fig. 1  Translational ($\Box$) and
orientational (+) correlations in harmonic lattices
with open boundaries, at finite temperature.  Upper set: $T=0.05$;
lower: $T=0.2$.  The straight lines represent
power-law fits to the translational correlation functions.
\vspace{1em}

\noindent Fig. 2. Snapshot of a frozen configuration in a
background potential of strength $S=0.1$, open boundaries.
\vspace{1em}

\noindent Fig. 3. Translational ($\Box$) and
orientational (+) correlations
in a harmonic lattice in the
presence of a background potential, $T=0$, open boundaries.
Upper set: $S=0.05$; lower: $S=0.5$.  The straight lines
represent exponential fits to the translational
correlation functions.
\vspace{1em}

\noindent Fig. 4. Translational ($\Box$) and
orientational (+) correlations
in a harmonic lattice in the
presence of a background potential, $T=0$, periodic boundaries.
Upper set: $S=0.05$; lower: $S=0.2$.  Solid lines
represent stretched exponential fits to the translational
correlation functions.

\end{document}